%% file: main.tex
\documentclass{article}

\usepackage[colorlinks,urlcolor=blue,linkcolor=blue,citecolor=blue]{hyperref}
\usepackage[subrefformat=parens,labelformat=parens]{subcaption} 
\usepackage{color,array}
\usepackage{graphicx}
\usepackage{cite}
\usepackage{amsmath,amssymb,amsfonts}
\usepackage{tabularx} 
\usepackage{booktabs}
\usepackage{multirow}
\usepackage{multicol}
\usepackage[table]{xcolor}
\usepackage[margin=1in]{geometry}
\begin{document}

\title{Evaluating Cooperative Resilience in Multiagent Systems: A Comparison Between Humans and LLMs}

\author{
    \small Manuela Chacon-Chamorro \and
    \small Juan Sebastián Pinzón \and
    \small Rubén Manrique \and
    \small Luis Felipe Giraldo \and
    \small Nicanor Quijano 
    \thanks{This work was supported by Google through the Google Research Scholar Program, and by the UniAndes–DeepMind Scholarship 2023.}
    \thanks{The authors are with the School of Engineering, Universidad de los Andes, Bogotá, Colombia (emails: \{m.chaconc, js.pinzonr, rf.manrique, lf.giraldo404, nquijano\}@uniandes.edu.co)}
    \thanks{This manuscript has been submitted to IEEE Transactions on Artificial Intelligence.}
}

\date{\small \today}


\maketitle
\hrule
\begin{abstract}
This paper presents a comparative analysis of cooperative resilience in multi-agent systems, defined as the ability to anticipate, resist, recover from, and transform to disruptive events that affect collective well-being. We focus on mixed-motive social dilemmas instantiated as a \textit{Tragedy of the Commons} environment from the Melting Pot suite, where we systematically compare human groups and Large Language Model (LLM)-based agents, each evaluated with and without explicit communication. Cooperative resilience is assessed under a continuously disruptive condition induced by a persistent unsustainable consumption bot, together with intermittent environmental shocks implemented as stochastic removal of shared resources across scenarios. This experimental design establishes a benchmark for cooperative resilience across agent architectures and interaction modalities, constituting a key step toward systematically comparing humans and LLM-based agents. Using this framework, we find that human groups with communication achieve the highest cooperative resilience compared to all other groups. Communication also improves the resilience of LLM agents, but their performance remains below human levels. Motivated by the performance of humans, we further examine a long-horizon setting with harsher environmental conditions, where humans sustain the shared resource and maintain high resilience in diverse disruption scenarios. Together, these results suggest that human decision-making under adverse social conditions can inform the design of artificial agents that promote prosocial and resilient behaviors.

\end{abstract}


\vspace{0.2 cm}
\hrule

\section{Introduction}
\input{sections/introduction}

\section{Framing Cooperative Resilience}
\input{sections/framing}

\section{Measuring Cooperative Resilience}
\input{sections/measuring}

\section{Analyzing Cooperative Resilience}
\input{sections/analyzing}

\section{Conclusions and Implications}
\input{sections/conclusions}

\section*{Ethics Statement}
All human-subject experiments were approved by the Institutional Review Board and Ethics Committee of the School of Engineering at Universidad de los Andes, Colombia (Approval ID~298, issued on July~18,~2025).





\bibliographystyle{IEEEtran}
\bibliography{references}

\end{document}

%% file: sections/introduction.tex

\label{sec:introduction}

The study of resilience in multiagent systems is essential, given that interactions in these settings typically occur under uncertain and dynamic conditions where external or internal disturbances threaten the stability of collective outcomes \cite{dafoe2020open, hammond2025multi, rizk2018decision}. Recent work introduced \textbf{Cooperative Resilience} as a system-level property describing the ability to maintain collective welfare despite disturbances, together with a quantitative framework to evaluate it in cooperative AI environments \cite{chacon2024cooperative}. In that work, the framework was validated with RL and LLM-based agents, showing that cooperative resilience complements traditional performance metrics by explicitly capturing adaptation and recovery under adversity \cite{chacon2024cooperative}. Closely related work on group resilience in multiagent Reinforcement Learning (RL) similarly shows that measuring resilience under perturbations is crucial for characterizing and designing robust collective behavior \cite{shraga2025collaboration}.

Building on this foundation, the present study expands the analysis by comparing human agents and LLM-based agents interacting in mixed-motive environments. This responds to a key gap in Cooperative AI: the lack of a benchmark for cooperative resilience that jointly evaluates humans and agents based on LLM under comparable conditions. We evaluate cooperative resilience in the social dilemma of the \textit{Tragedy of the Commons} \cite{perolat2017multi,janssen2010lab}, using the Melting Pot 2.0 simulation suite \cite{agapiou2022melting}, comparing groups composed of humans and LLM agents, both with and without communication capabilities. Communication was selected as a key skill to evaluate, as previous studies in multiagent systems have highlighted its central role in enabling coordination, fostering cooperation, and improving collective performance \cite{dafoe2020open, havrylov2017emergence, balliet2010communication, kim2019learning}. To assess resilience, we introduce a permanent disruption via an unsustainable RL agent (the \emph{disruptive bot}), whose individual strategy undermines collective welfare, and add environmental shocks through intermittent removal of shared resources. Each group (humans or LLMs, with or without communication) is evaluated in nine scenarios that vary the probability and frequency of resource elimination.

Comparing human groups with agents based on LLM in social dilemmas offers an opportunity to examine how different forms of decision-making shape cooperative resilience \cite{chuang2023wisdom, willis2025will}. Human agents bring adaptive strategies, social heuristics \cite{rand2014social}, and the ability to leverage communication in creative ways \cite{kopp2021revisiting}. Moreover, human communication involves incremental co-construction and \textit{mentalization}, enabling  coordination far beyond predefined protocols \cite{kopp2021revisiting}. In contrast, LLM agents rely on predefined language-based reasoning and limited experiential grounding. Recent studies have begun to examine their cooperative ability and performance in social settings, highlighting both promising behaviors and fundamental limitations \cite{mosquera2025can, liu2024exploring, zhu2025learning, shraga2025collaboration, smith2025evaluating}.

Our results reveal significant differences between humans and LLM agents. Human groups with communication exhibit higher cooperative resilience than all other conditions, suggesting that communication and human decision-making strategies play a central role in sustaining collective welfare under adversity. LLM agents benefit from communication; however, their resilience does not reach the level displayed by humans. These findings underscore the importance of incorporating human-inspired mechanisms when designing artificial agents that aim to promote prosocial and resilient behaviors. With this work, the experimental analysis presented in \cite{chacon2024cooperative} is expanded, establishing a comparative line that is expected to extend to hybrid groups and the integration of human-inspired insights into the design of future cooperative AI systems.

The remainder of this paper is organized as follows. Section~\ref{sec:measuring} introduces the experimental setup and the methodology for measuring cooperative resilience; Section~\ref{sec:analyzing} presents and discusses the results; and Section~\ref{sec:conclusions} offers concluding remarks and outlines future directions.

\begin{figure*}[t]
\footnotesize
  \centering
  \includegraphics[width=\textwidth]{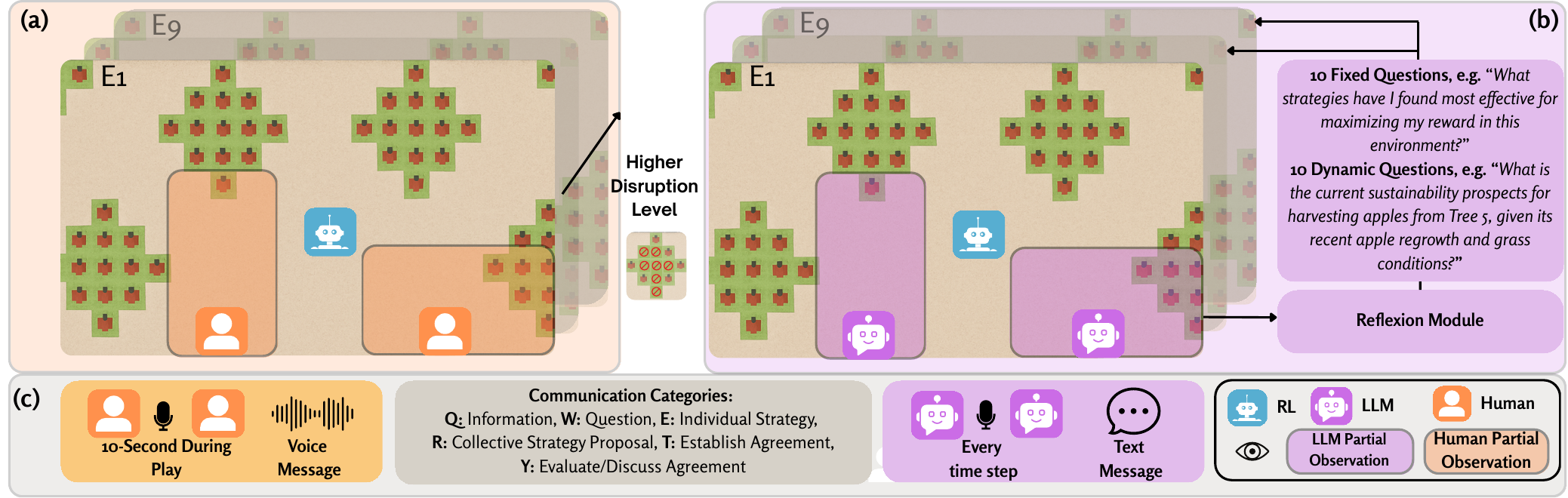}
  \caption{\footnotesize Experimental setup for cooperative resilience comparing human and LLM agents, with and without communication. \textbf{(a)} Human setup: sequential curriculum E1$\to$E9. \textbf{(b)} LLM setup: same curriculum plus a reflection module (10 fixed + 10 dynamical generated questions) that accumulates persistent memories across episodes. \textbf{(c)} Communication protocol: humans exchange 10-s voice messages during play; LLMs emit text each timestep; both use six categories.}
  \label{fig:experiment_scenario}
\end{figure*}

%% file: sections/framing.tex
Resilience is broadly understood as the ability of a system to sustain desirable functioning under change or disruption \cite{2_1, 3_3}. In engineering, it refers to the capacity to anticipate, absorb, recover from, and adapt to perturbations, going beyond robustness, which merely resists change without reorganization \cite{kouicem2020towards}. This system-level view has inspired recent work in multiagent settings, where agents must preserve cooperation when incentives or environmental conditions shift \cite{chacon2024cooperative, shraga2025collaboration}, and where communication and information sharing are critical for maintaining coordination and restoring collective performance after disturbances \cite{abu2021promoting, dafoe2020open, du2025contextual, shraga2025collaboration}.

In parallel, human-centered studies show that resilience in collective behavior hinges on social mechanisms such as trust, reciprocity, and reputation: communication and identifiability promote cooperation, while adaptive responses like turning or sanctioning help sustain it over time \cite{hughes2025modeling}. Work on human–agent teams further indicates that the cooperative design of artificial partners shapes trust and recovery after unexpected events \cite{1_4}. More recently, LLMs have been studied as social agents capable of reasoning, negotiation, and coordination through language \cite{zhu2025learning, mosquera2025can, liu2024exploring, tran2025multi}; however, \textbf{no consistent benchmark has yet been established that compares their cooperative resilience to that of humans under disruptive conditions.} 

Building on these perspectives, we frame cooperative resilience as a measurable property that captures the ability of agents, human or artificial, to sustain collective welfare under disruptions, with communication serving both as a coordination medium. Our comparative study is targeted at a controlled analysis in mixed-motive settings, but naturally points to two broader implications: (i) a compact initial benchmark against which subsequent agent designs can be evaluated—and ideally surpassed; and (ii) a basis for extracting some design-relevant insights about the skills and behaviors that underpin cooperative resilience.

%% file: sections/measuring.tex
\label{sec:measuring}

\subsection{Experiment setup}

The experiments were conducted using the Melting Pot 2.0 simulation suite \cite{agapiou2022melting}, a multiagent environment specifically designed to study social dilemmas and mixed-motive interactions. The selected scenario corresponds to the \textit{Commons Harvest}, where agents harvest apples that regenerate depending on collective consumption \cite{perolat2017multi}. To allow direct comparison between groups, the environment was adapted for both human participants and agents based on LLM, ensuring that the interface, rules, and available actions were consistent across conditions. Figure ~\ref{fig:experiment_scenario} summarizes the experimental setup for humans and LLM agents.

For human participants, the Melting Pot environment was adapted to allow remote desktop access, allowing each individual to control their own agent through a custom interface. The interface allowed participants to perform movement and attack actions using the keyboard. Before each session, participants were instructed on the individual objective: \textit{to collect as many apples as possible from the six trees distributed throughout the environment}. They were also informed that \textit{apples would regenerate over time depending on the number of apples remaining on each tree}. \textit{If the last apple from a tree was taken, the tree would disappear and no longer regenerate.} Each player had a partial observation of the environment, consistent with the observation model used by artificial agents in Melting Pot. This ensured that human participants experienced the same local first-person perspective as agents.

For LLM agents, we implement an observation-to-text adapter that renders each agent's spatial observation into inputs suitable for GPT-4. The pipeline has two stages: (i) a raster-to-ASCII serializer that encodes the agent-centric field of view in real time, and (ii) a semantic summarizer that converts this ASCII map into a concise natural-language description of salient entities, relations, and affordances. This design follows recent trends in LLM-based multiagent systems \cite{cross2024hypothetical} and is integrated into the agent's decision loop using the Generative Agents architecture \cite{mosquera2025can} enabling text-conditioned action selection. 

The scenarios used to evaluate cooperative resilience consist of nine distinct conditions, derived from a combination of two variables: (i) the number of disruption events applied during the episode ($d = 1, 2, 3$) and (ii) the magnitude of the disruption, represented by the probability ($v_s = 0.3, 0.5, 0.7$) of removing individual apples. During each disruption event, the environment iterates over all trees containing more than one apple. For each apple on such trees, a Bernoulli trial with probability $v_s$ determines whether the apple is removed. This selective removal simulates resource degradation while preserving trees with scarce availability.

For humans, we used a sequential curriculum from E1 ($v_s=0.3$, one disruption) to E9 (three disruptions, highest removal probability) to promote learning with increasing difficulty. In addition, a continuously disruptive agent with an unsustainable harvesting strategy collects apples persistently throughout each episode. To ensure comparability with the human sequence, LLM agents also follow a sequential curriculum from E1 to E9. After each scenario, the agent runs a reflection pipeline with 20 prompts: 10 fixed, generic questions, and 10 dynamic, scenario-specific questions that are automatically generated from the agent's current interaction logs (i.e., not predefined). The answers to these questions are used to extract and filter memories that capture durable and transferable insights. These \textit{long-lived insights} are attached to a persistent memory bank and carried forward from E1 through E9. Further details, including question templates and LLM architecture specifications, are provided in the Supplementary File.

The experiments were carried out under two conditions for each group: with and without communication. In the communication-enabled setting, humans participants were allowed to interact using natural language voice communication while playing. To structure the communication, we restricted communication to six categories, each bound to a hotkey: \textbf{Q} (Information), \textbf{W} (Question), \textbf{E} (Individual Strategy), \textbf{R} (Collective Strategy Proposal), \textbf{T} (Establish Agreement), \textbf{Y} (Evaluate/Discuss Agreement). 

In the LLM condition, communication occurs at every timestep via a structured, turn-aware channel. At each turn, an agent first decides whether to speak based on its current state and observation, the chosen action, recent partner messages, active goals/plans, and any standing agreements. When communicating, the agent emits one of six message types subject to a 50-word cap to promote concise, realistic exchanges. The channel enforces sequential visibility: the first agent in a round only observes messages from the previous round, whereas later agents also observe current-round messages already sent, approximating real-time ordered interaction. All messages are written in long-term memory, enabling agents to track others' statements, their own commitments, and to build persistent coordination patterns over episodes. (Details are provided in the Supplementary File, Sec. 1.3.)

\subsection{Indicators and Quantification of Cooperative Resilience}

We adopt the methodology from \cite{chacon2024cooperative} to compute a cooperative resilience score $\rho$. A score of $\rho = 1$ indicates alignment with a baseline without disruption, values below 1 reflect loss of resilience, and scores above 1 suggest exceptional recovery or improved performance after disruptions. This score aggregates four indicators of collective well-being: (i) Apple Availability, measured as the number of apples present in the grid at each step; (ii) Tree Preservation, measured as the number of intact trees, which reflects the extent to which agents avoid short-sighted, selfish harvesting that depletes future resources; (iii) an Equality Gini Index over consumption, capturing how evenly resources are distributed; and (iv) a Satiation, estimating the delay between successive resource accesses per agent. The formal definitions and equations for these indicators are provided in the Supplementary File (Section 2).

Each indicator is measured under two conditions: a \textit{baseline} scenario with no disruption and a \textit{disrupted} scenario. Let $I_k(t)$ denote the time series of indicator $k$ over time. Let $t_d$ be the disruption time, $t_f$ the worst degradation time, and $t_r$ the recovery endpoint (typically the end of the episode).

We define the failure $FP$ and recovery $RP$ profiles as follows.

\begin{equation*}
\begin{minipage}{0.45\linewidth}
\[
\text{FP}_k = 
\int_{t_d}^{t_f} \frac{I_k^{\text{disrupted}}(t)}{I_k^{\text{baseline}}(t)} \, dt
\]
\end{minipage}
\hfill
\begin{minipage}{0.45\linewidth}
\[
\text{RP}_k =
\int_{t_f}^{t_r} \frac{I_k^{\text{disrupted}}(t)}{I_k^{\text{baseline}}(t)} \, dt
\]
\end{minipage}
\end{equation*}

We denote the durations: $\Delta t_f = t_f - t_d, \qquad \Delta t_r = t_r - t_f$, and compute the resilience score for the indicator $k$ as:

\begin{equation*}
\rho_k = \frac{t_d + \text{FP}_k \cdot \Delta t_f + \text{RP}_k \cdot \Delta t_r}{t_d + \Delta t_f + \Delta t_r}.
\end{equation*}

Finally, we aggregate indicator-specific resilience scores into a global resilience score using the harmonic mean.

\begin{equation*}
\rho(\tau) = \left( \frac{1}{K} \sum_{k=1}^{K} \frac{1}{\rho_k} \right)^{-1}.
\end{equation*}

This methodology quantifies cooperative resilience by contrasting system behavior with and without disruption, extracting failure and recovery profiles for each collective-welfare indicator, and aggregating them through a harmonic mean \cite{chacon2024cooperative}. In doing so, it provides a single interpretable score that summarizes how well the system withstands, recovers from, and adapts to adverse conditions.

\subsection{Shared Baseline Construction}

In the original methodology proposed in \cite{chacon2024cooperative}, cooperative resilience was assessed by comparing the performance of the system in the presence of disruptions with its own behavior in the absence of these disruptions. This approach is valid when the goal is to measure cooperative resilience as an intrinsic property of a system, evaluating this ability relative to its own undisturbed baseline.

However, the current study aims to compare different types of agents and the presence or absence of communication. In this setting, each group exhibits distinct baseline behaviors even in the absence of disruptions, which challenges the comparability of cooperative resilience scores computed relative to heterogeneous baselines. To enable fair and cross-group comparisons, we adapt the original methodology by incorporating common baseline references shared between all conditions.

For each experimental group, humans and LLMs, with and without communication, we first computed the mean trajectory across five episodes for each metric. Then, we calculate the temporal median of each group's average curve, obtaining a robust summary of its behavior. Finally, we define a common baseline for the metric as the median of these four group-specific medians, ensuring a single, fair, and comparable reference point across all conditions (see the Supplementary File Section 3). 

%% file: sections/analyzing.tex
\label{sec:analyzing}

Using the methodology described above, we evaluated the cooperative resilience of humans and agents based on LLM under different conditions. This metric captures how well each group maintains collective welfare under disruption,  how much performance degrades during adverse events, and how quickly it recovers afterward.

\subsection{Short-horizon comparison: Humans vs LLMs}

\begin{figure}[t!]
\footnotesize
     \centering
     \begin{subfigure}{0.34\textwidth}
         \centering
        \includegraphics[width=\textwidth]{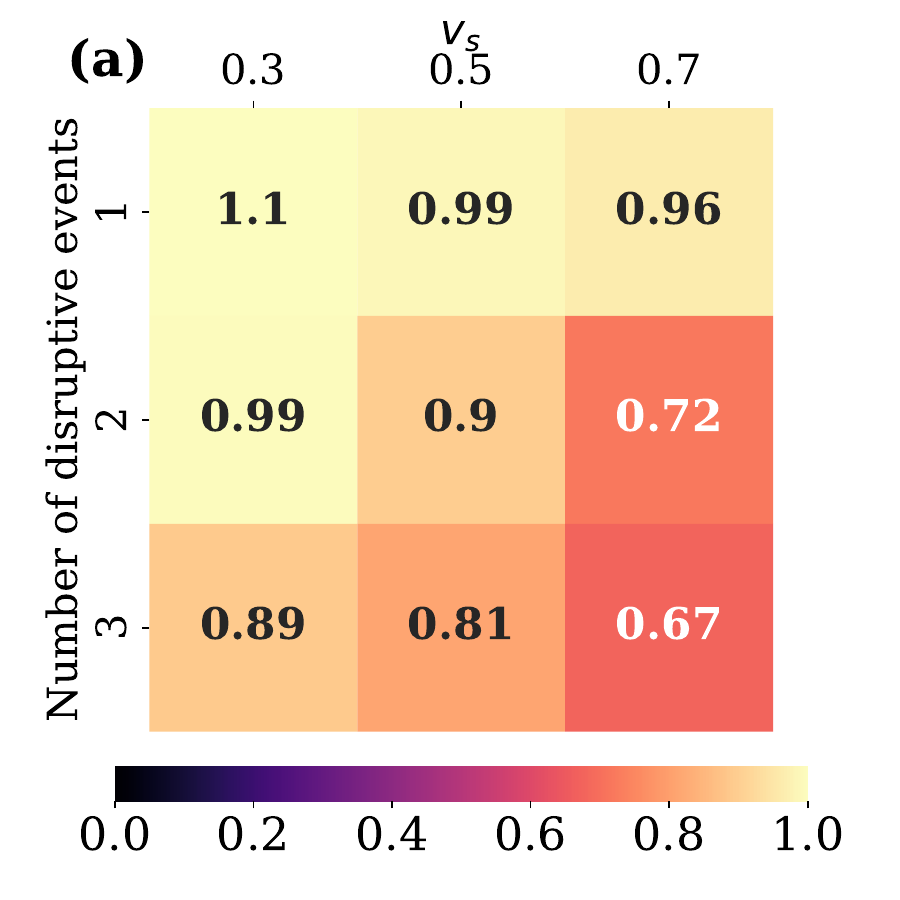}
     \end{subfigure}
     \begin{subfigure}{0.34\textwidth}
         \centering
        \includegraphics[width=\textwidth]{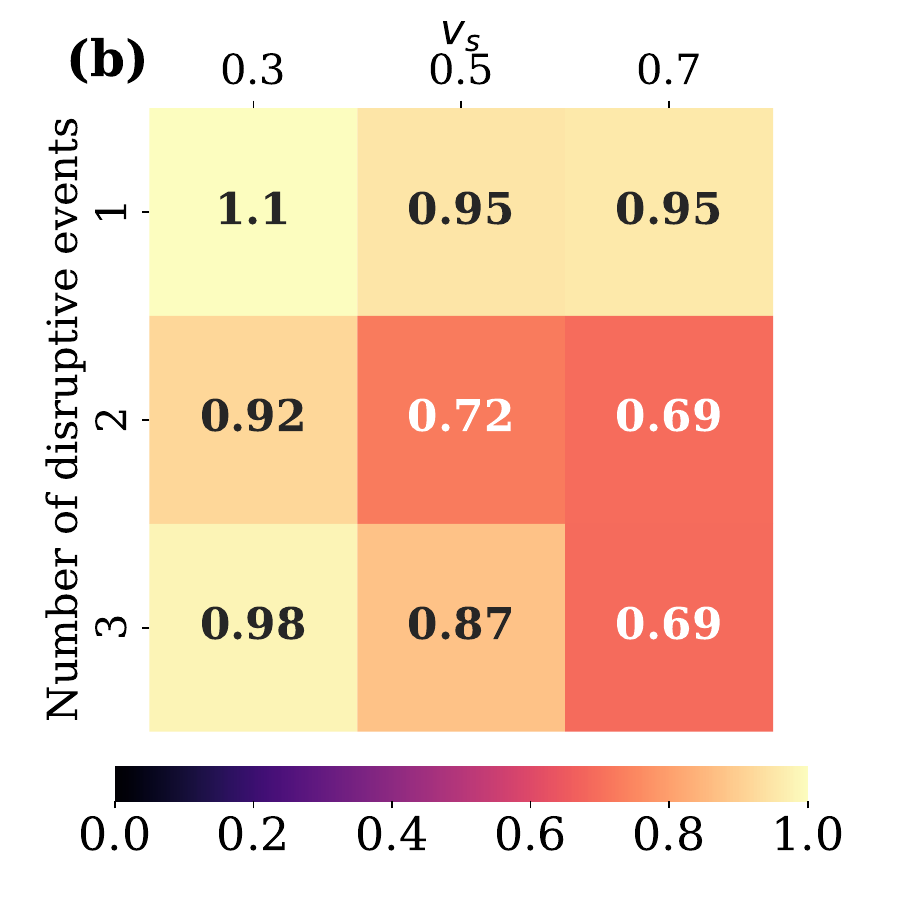}
     \end{subfigure} 
     
     \begin{subfigure}[b]{0.34\textwidth}
     \vspace{-0.3cm}
        \centering
        \includegraphics[width=\textwidth]{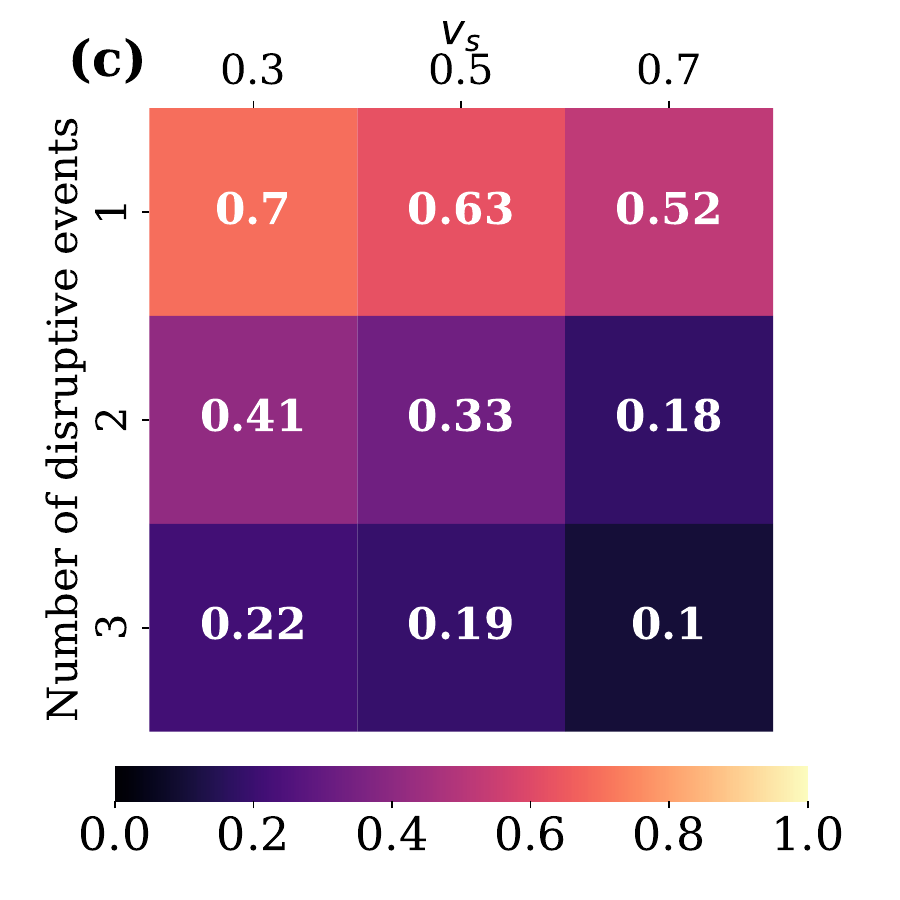}
     \end{subfigure}
     \begin{subfigure}[b]{0.34\textwidth}
         \centering
         \includegraphics[width=\textwidth]{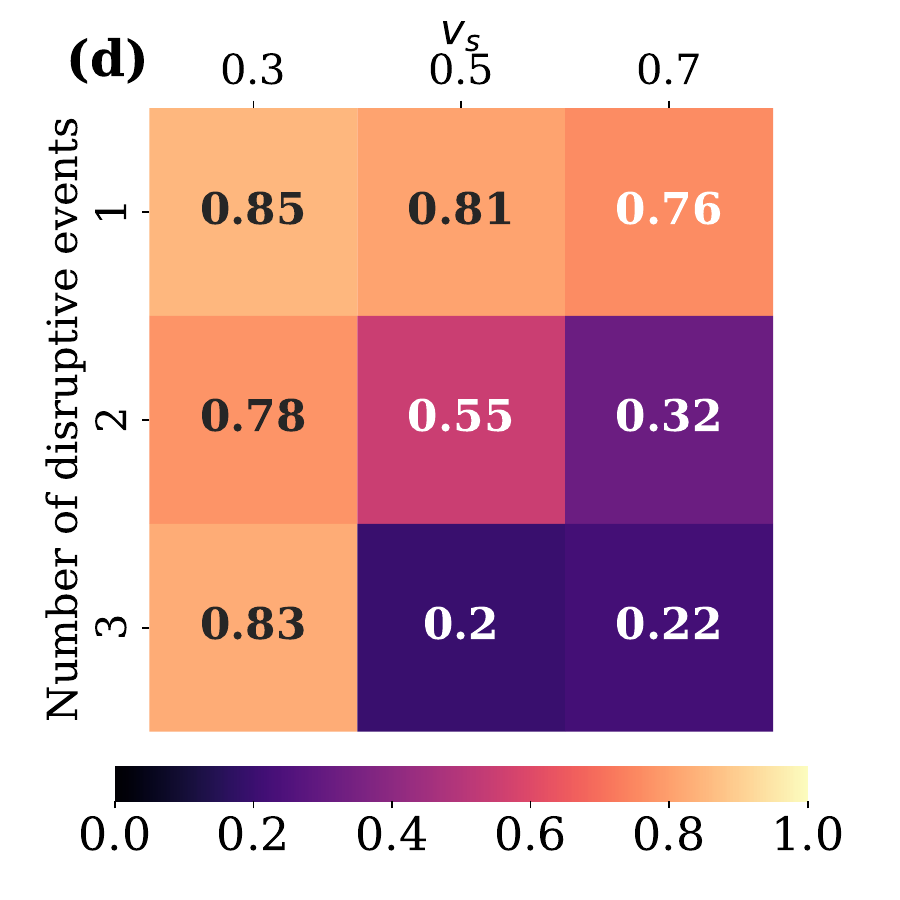}
     \end{subfigure}
    \caption{\footnotesize Cooperative resilience values across nine scenarios. a) Humans without communication, b) Humans with communication, c) LLMs without communication, and d) LLMs with communication. Results are presented as heatmaps, where rows represent the number of disruptive events (apple removals) and columns indicate the probability ($v_s$) of resource elimination. These two dimensions are directly related to system vulnerability and risk. Color intensity encodes the level of cooperative resilience, with lighter tones indicating higher resilience values and darker tones reflecting lower resilience.}
     \label{fig:resilienceMaps}
\end{figure}

To synthesize the nine experimental scenarios, the results for each group are summarized in heatmaps presented in Figure ~\ref{fig:resilienceMaps}, which capture resilience values as a function of two factors: the number of disruptive events and the probability of resource elimination ($v_s$). Each map corresponds to one group configuration (humans with communication, humans without communication, LLMs with communication, and LLMs without communication). The intensity of the color encodes the level of cooperative resilience, with lighter shades representing higher values. This visualization allows for direct comparison of conditions, highlighting both the sensitivity of each group to disruptions and the role of communication in sustaining collective welfare.

Across all disruption settings, human groups exhibit consistently higher cooperative resilience than LLM-based agents. Panels (a) and (b) show that, for humans, the presence of communication yields only a modest improvement: resilience remains high and remarkably stable even as disruptions intensify, from one to three disruptive events and from low to high resource, elimination probability. This stability highlights a strong intrinsic ability to adapt, preserve coordination, and prevent collective collapse under increasingly adverse conditions, regardless of whether explicit communication is available.

In contrast, LLM-based agents show a markedly different pattern. Without communication (panel c), resilience deteriorates sharply as scenarios become more demanding. This degradation indicates limited adaptive capacity: although the agents carry a memory module across episodes, they do not leverage past experience to maintain robust behavior as disruptions escalate. Each increase in scenario difficulty results in a substantial performance drop, revealing sensitivity to perturbations and a lack of compensatory strategic adjustment. 

When communication is enabled (panel d), resilience improves across all disruption levels. This suggests that explicit message exchange is a crucial coordination mechanism for LLM agents, supporting recovery collective behavior in the presence of shocks. Communication does not prevent resilience from dropping in the most severe scenarios, but it reshapes the sensitivity profile: instead of a uniform collapse as disruptions intensify, LLM groups preserve high resilience in several intermediate conditions and only exhibit strong degradation when both the number of disruptive events and $v_s$ are simultaneously high.

In general, the results show a clear behavioral separation between humans and LLM-based agents. Human groups preserve high resilience even under harsher disruptions, a pattern consistent with robust intrinsic cooperative capabilities such as flexible strategic adjustment, norm sensitivity, and real-time coordination. In contrast, LLM agents display high sensitivity to perturbations and are highly dependent on communication to maintain stability between scenarios. This indicates that explicit communication channels are essential for enabling scalable cooperative behavior in current LLM-based agents, particularly in mixed-motive settings or environments where disruptions accumulate over time.

\begin{figure}[t]
\footnotesize
    \centering
    \includegraphics[width=0.7\linewidth]{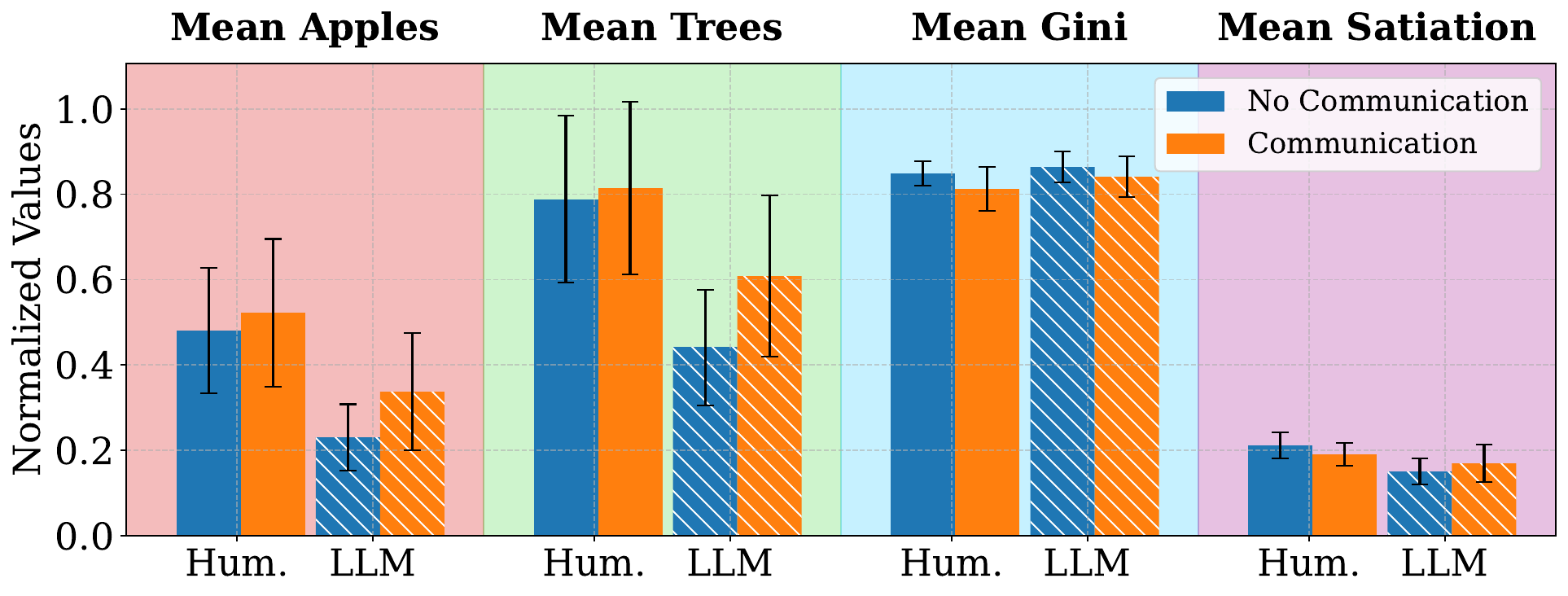}
    \caption{\footnotesize Mean $\pm$ standard deviation of environmental and social metrics across the nine scenarios for both human conditions, with and without communication. Apple and tree values are normalized by their maximum possible quantities (64 apples and 6 trees, respectively).}
    \label{fig:human_LLM_short_metrics}
\end{figure}

To complement this analysis, we examine environmental and social metrics. To assess environmental availability, a proxy for sustainability, in Figure ~\ref{fig:human_LLM_short_metrics} we plot bar charts of the mean (± standard deviation) of apples and trees across all scenarios and groups, normalized by their maximum capacities (64 apples and 6 trees). We also report episode-averaged equality, a sustained-fairness proxy, and the satiation index. In Figure ~\ref{fig:human_LLM_short_metrics} human groups leave more apples and trees in the environment than LLM agents, with or without communication, reinforcing their horizon sustainability. Communication has only a marginal (and statistically unclear) effect on humans but noticeably increases the remaining apples and trees of LLM agents, although not up to human levels. The equality index is comparable across populations, yet it slightly decreases under communication, likely because agents restrict the unsustainable bot's access to resources. The satiation index is also similar for humans and LLMs, suggesting that both populations access resources at comparable rates; a plausible explanation for the sustainability gap is that humans refrain from consuming the last available apples, whereas LLM agents tend to exhaust them. However, these aggregate indicators only provide static summaries of outcomes and are less sensitive to how systems degrade and recover over time, which underscores the role of the cooperative resilience metric in capturing adaptation and recovery dynamics along the full trajectory.

\subsection{Long-horizon sustainability: Human benchmark}

The previous results show that humans, regardless of communication condition, achieve higher resilience and typically finish episodes with non-depleted resources (see Figures S3-S4 in the Supplementary File). By contrast, LLM agents tend to exhaust all resources by the end of the simulation (see Figures S7-S8 in the Supplementary File). This pattern motivates the hypothesis that humans exhibit greater sustainability, a property closely tied to cooperative resilience. Consequently, we design a new experimental series with a longer horizon (1000 steps; 4-times the original) and a lower apple regrowth rate to simulate a more resource-constrained environment. We conducted these experiments only with humans, since the LLM groups systematically exhausted the shared resources, causing the environment to collapse before 250 time steps. As before, we evaluate the nine scenarios (E1–E9) sequentially, increasing the severity of the disturbance at each step.

\begin{figure}[t]
\footnotesize
     \centering
     \begin{subfigure}[b]{0.34\textwidth}
         \centering
        \includegraphics[width=\textwidth]{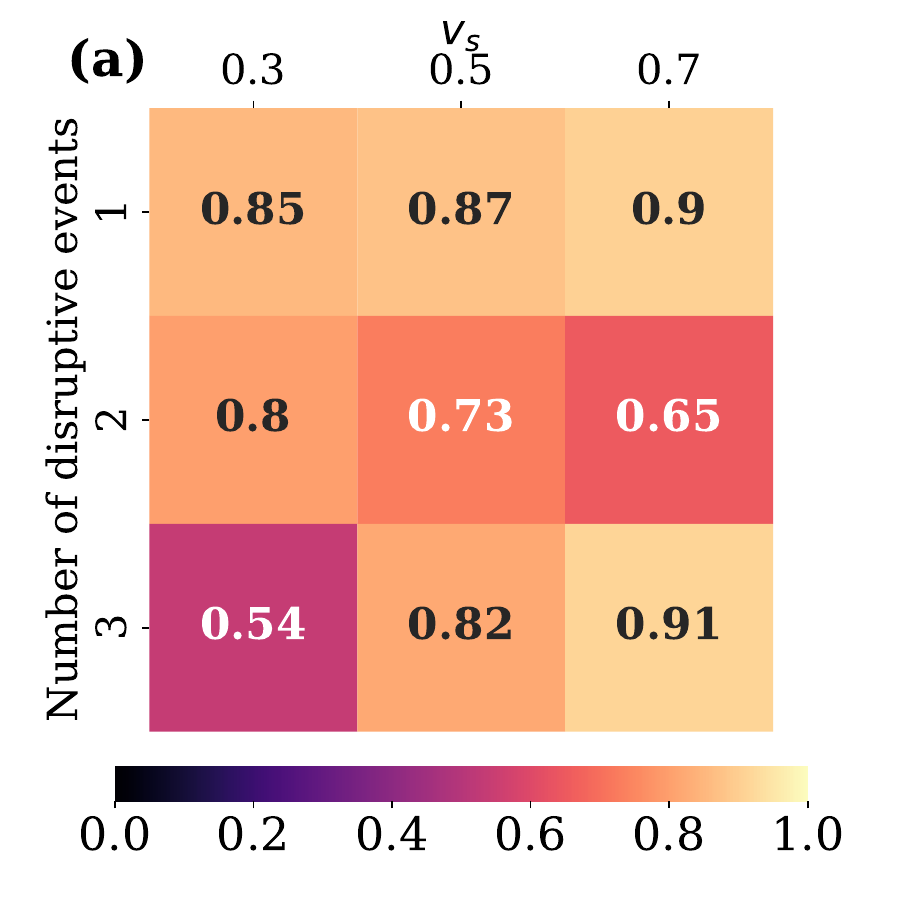}
     \end{subfigure}
     \begin{subfigure}[b]{0.34\textwidth}
         \centering
        \includegraphics[width=\textwidth]{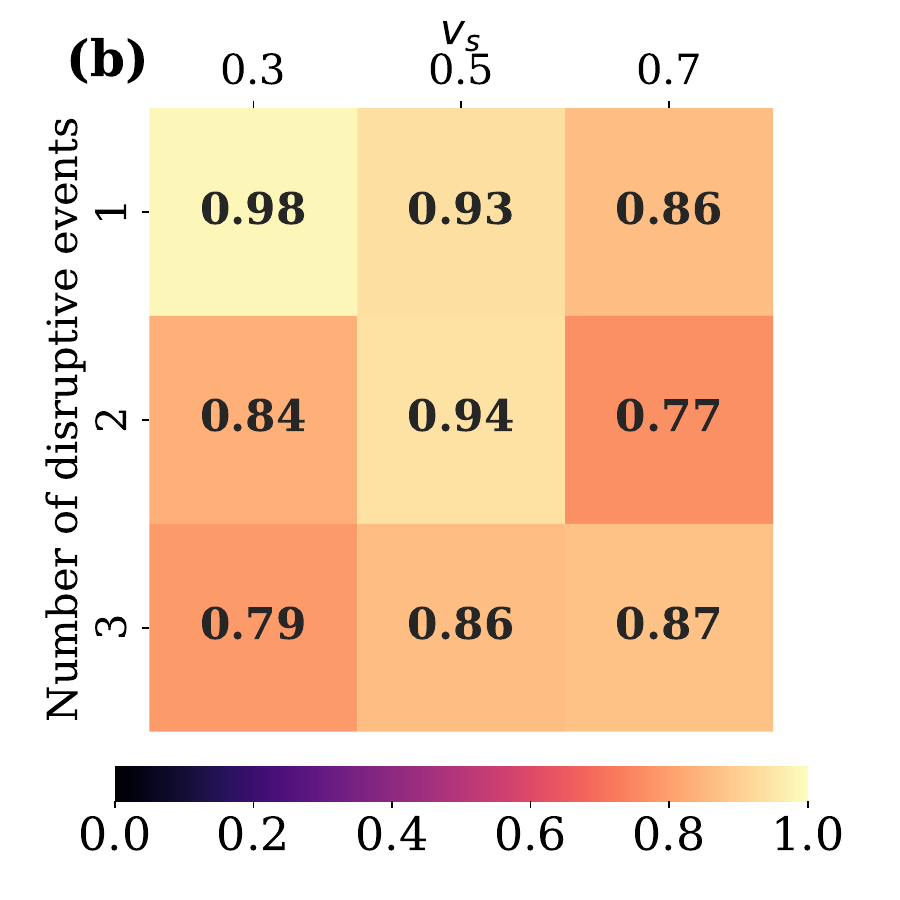}
     \end{subfigure} 
         
    \caption{\footnotesize Cooperative resilience values across nine scenarios. a) Humans without communication, b) Humans with communication. Rows represent the number of disruptive events (apple removals) and columns indicate the probability ($v_s$) of resource elimination. These two dimensions are directly related to system vulnerability and risk. Color intensity encodes the level of cooperative resilience, with lighter tones indicating higher resilience values and darker tones reflecting lower resilience.}
     \label{fig:resilienceMaps_longer}
\end{figure}

Figure ~\ref{fig:resilienceMaps_longer} reports cooperative resilience in the extended 1000-timestep environment. Humans maintain high and stable resilience under almost all disruption conditions, despite the lower regrowth rate and the long horizon. Panel (a) (no communication) already displays strong performance, with only one noticeable drop in the mild-disruption setting ($v_s=0.3$, three events). In addition, resilience rises again in harsher scenarios, suggesting that participants adapt their strategies as the scenarios progress. 

In panel (b) (communication), resilience becomes more uniform across scenarios: all values exceed 0.79 and show reduced sensitivity to the joint increase in disruptive events and removal probability. Although not every individual cell surpasses its noncommunication counterpart, the gains are most pronounced in the harsher disruption settings, indicating that communication primarily strengthens coordination and robustness when the environment is more demanding.

\begin{figure}[b]
\footnotesize
    \centering
    \includegraphics[width=0.6\linewidth]{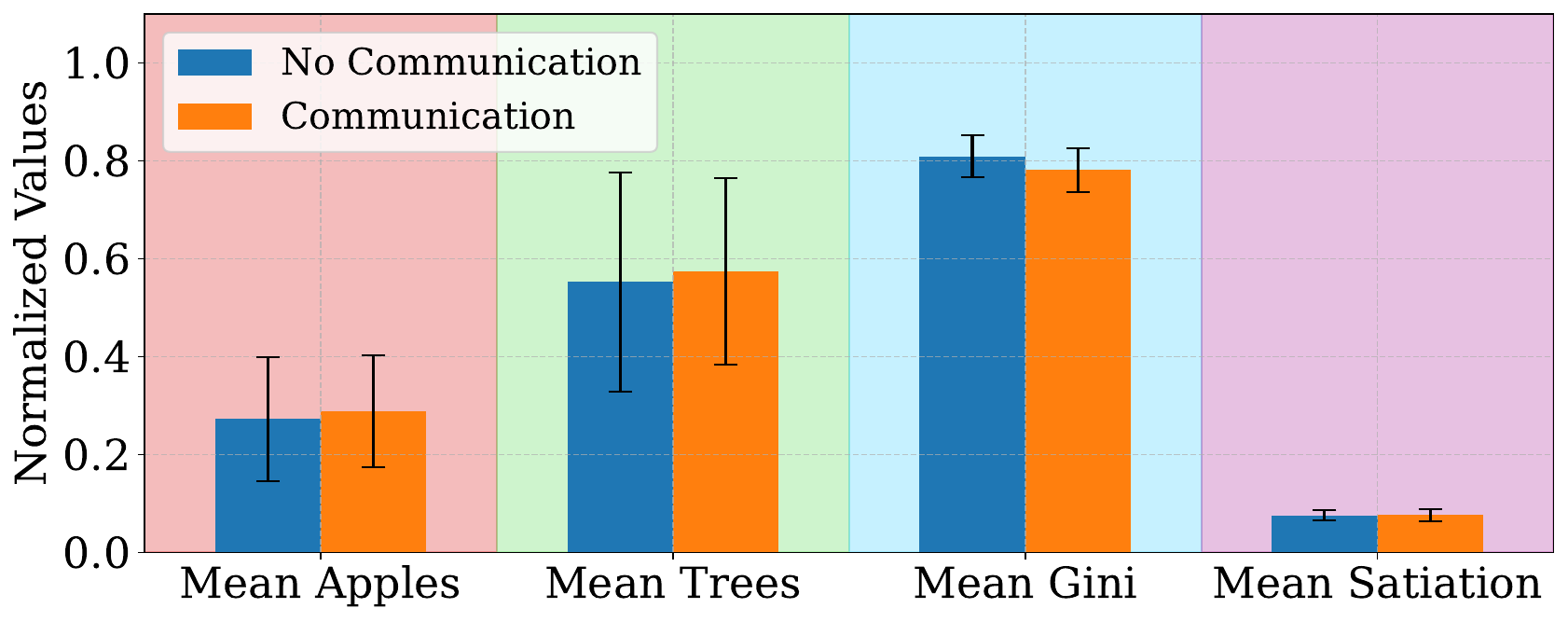}
    \caption{\footnotesize Mean $\pm$ standard deviation of environmental and social metrics across the nine scenarios for both human conditions, with and without communication. Apple and tree values are normalized by their maximum possible quantities (64 apples and 6 trees, respectively).}
    \label{fig:human_long_metrics}
\end{figure}

Figure ~\ref{fig:human_long_metrics} presents the mean values of remaining apples and trees in the environment, the equality gini index and the satiation index, computed across the nine scenarios for both human conditions, with and without communication. The bar chart shows that, in the long run, human groups display similar aggregate outcomes with and without communication. In the figure apple availability decreases to roughly approx. 25 \% of its initial capacity, but trees stabilize around 60 \%, indicating that participants tend to avoid overharvesting and preserve at least one apple per tree, a behavior consistent with sustainable reasoning. Equality remains high in both cases, although slightly lower with communication, possibly because participants collectively prevent the bot agent from monopolizing resources, identifying it as a source of unsustainable behavior. The satiation index remains high and comparable across conditions, though it is slightly lower with communication, possibly because participants strategically restrict the bot's access to resources. These results suggest that communication only marginally affects aggregate environmental metrics, but its impact becomes more evident in the cooperative resilience analysis. 


\subsection{Communication analysis across settings}

\begin{figure}[t]
\footnotesize
    \centering
    \includegraphics[width=0.6\linewidth]{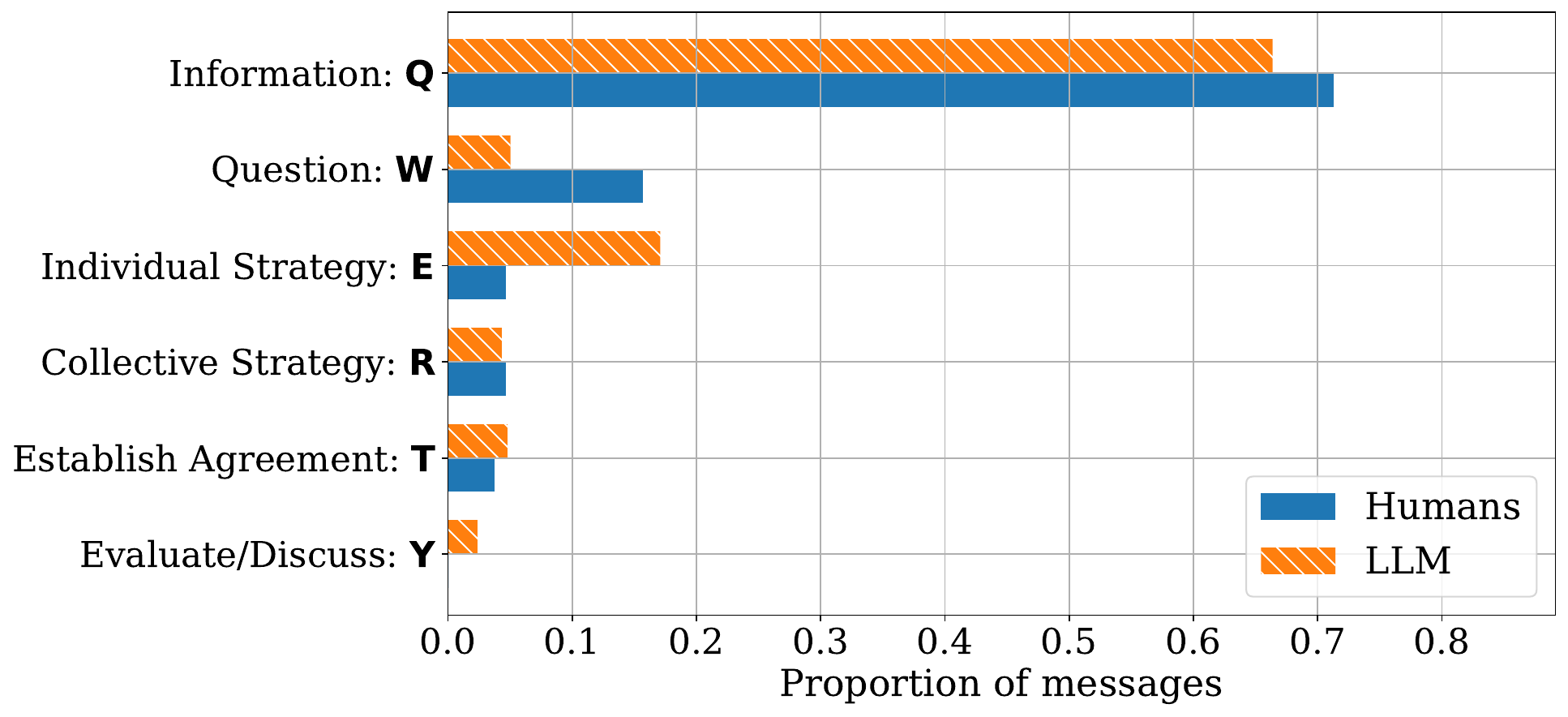}
    \caption{\footnotesize Proportion of message types exchanged across the nine resilience scenarios between human and LLM agents.}
    \label{fig:commnunication_metrics}
\end{figure}

In addition, we analyze the communication patterns exchanged during the experiments.  Figure ~\ref{fig:commnunication_metrics} presents the proportion of communication messages exchanged during the short human–agent and LLM–agent experiments. Both groups concentrate most of their communication in the \textbf{Q} category, indicating that the dominant function of messages, across humans and LLMs, is to share environment-relevant observations. Humans exhibit an even stronger bias toward information sharing, while the remaining categories appear with substantially lower frequency. A second notable divergence arises in the \textbf{W} and \textbf{E} categories: humans tend to ask more questions, whereas LLM agents allocate a larger share of their communication to describing their own planned actions. Interestingly, categories more closely tied to explicit coordination, \textbf{R}, \textbf{T}, and \textbf{Y}, remain marginal for both groups. This suggests that neither humans nor current LLM agents naturally adopt communication protocols geared toward forming and validating collective commitments. The limited presence of these messages is consistent with the larger challenges identified in the literature, where the establishment of shared plans, the negotiation of joint commitments and the maintenance of cooperative norms remain open problems for artificial agents \cite{dafoe2020open}, and also for humans operating in social dilemmas.

\begin{figure}[h]
\footnotesize
    \centering
    \includegraphics[width=0.45\linewidth]{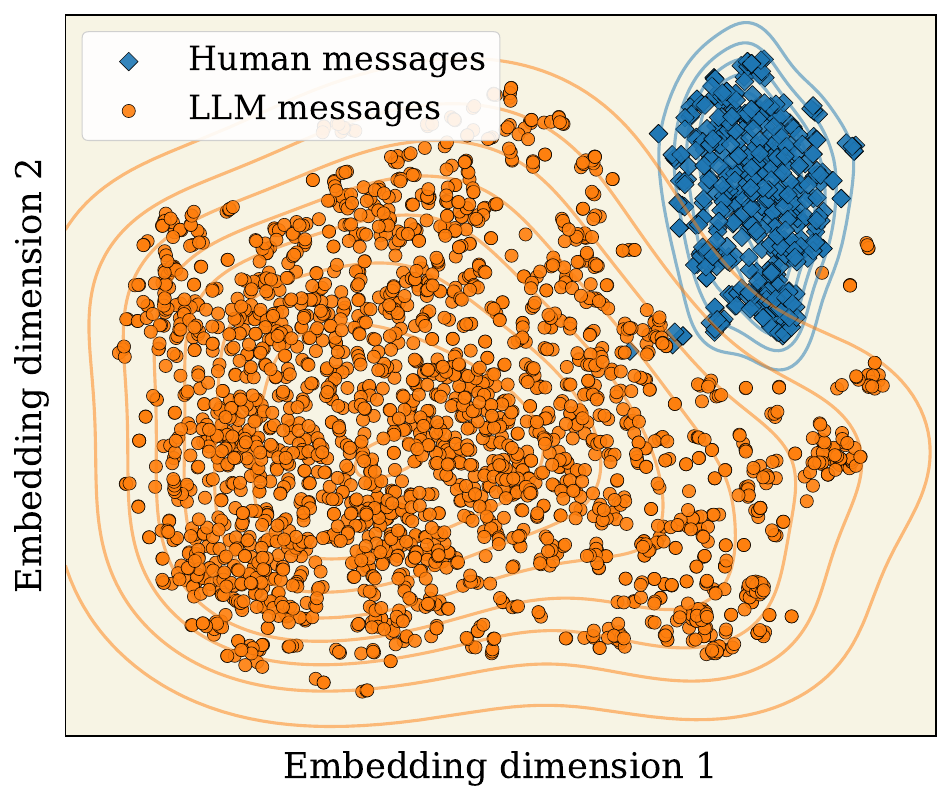}
    \caption{\footnotesize Semantic embedding space of human and LLM messages (t-SNE) with kernel-density contours. Each point is a message embedded with a multilingual encoder; contours show message density for each group.}
    \label{fig:embedding}
\end{figure}

To examine the semantic structure of communication, we embedded every message using the multilingual encoder distiluse-base-multilingual-cased-v2, which maps semantically equivalent content across languages into a shared vector space; given that human messages were produced in Spanish and LLM messages in English. The 2D t-SNE projection in Figure ~\ref{fig:embedding} reveals a clear separation between the two groups: human and LLM communications occupy distinct semantic regions. Human messages form a tight cluster, reflecting limited lexical variability, whereas LLM messages spread over a large, heterogeneous cloud, indicating higher lexical and syntactic diversity. The absence of overlap between these regions shows that humans and LLMs rely on fundamentally different semantic structures when communicating, suggesting that they do not converge toward a shared communicative manifold. This structural misalignment in how information is framed and grounded by artificial agents compared to human players may hinder the emergence of shared protocols for coordination in cooperative tasks.

\textbf{Summary:} Across short and long horizon experiments, and under multiple disruptive regimes, a consistent pattern emerges: human groups maintain higher cooperative resilience and greater environmental sustainability than LLM-based agents. Communication helps both populations, but humans benefit less because they already coordinate effectively, whereas LLM agents improve but remain below human levels. Finally, the semantic structure of the communication reveals distinct coordination strategies between humans and LLM agents. Together, these results provide a coherent picture of the behavioral gap that the cooperative-resilience benchmark is designed to expose.

%% file: sections/conclusions.tex

\label{sec:conclusions}

This work introduces a benchmark for evaluating cooperative resilience in mixed-motive environments, enabling systematic comparison of humans and agents based on LLM under controlled disruptive conditions. Through a measurement framework and a matched experimental design, the study demonstrates how resilience can be assessed through failure and recovery dynamics, beyond static performance indicators. Our results reveal a clear behavioral gap between humans and current LLM-based agents, emphasizing the need for improved cooperative reasoning, grounding, and communication strategies in artificial agents. Although our experiments focus on a specific LLM-based instantiation (GPT-4 with a generative agents architecture), the observed limitations are not necessarily model specific but may characterize a broader class of LLM agents relying on planning, reflection, and memory. The proposed benchmark provides a reproducible foundation for evaluating such advances and guiding the development of agent architectures, reward designs, and interaction protocols that support resilient and prosocial behavior. Future work should extend this framework to diverse environments, larger populations, and hybrid human–LLM teams, paving the way for artificial agents capable of sustaining cooperation and collective welfare under adverse conditions.